\documentclass{elsart}
\journal{Physics Letters B}

\usepackage{natbib,amssymb,graphicx}
\usepackage{pstricks} 

\newcommand{\pT}{p_\perp}

\newcommand{\Pythia}{\textsc{Pythia}}
\newcommand{\Jetset}{\textsc{Jetset}}
\newcommand{\Fritiof}{\textsc{Fritiof}}

\newcommand{\UNIT}[1]{\mbox{$\,{\rm #1}$}}

\newcommand{\GeV}{\UNIT{GeV}}

\newcommand{\fm}{\UNIT{fm}}

\newcommand{\fmc}{\UNIT{fm/c}}

\newcommand{\REM}[1]{}

\begin{document}

\begin{frontmatter}
\title{Space-Time Picture of Fragmentation in PYTHIA/JETSET for HERMES and RHIC} 
\author{K.~Gallmeister\corauthref{cor}},
\corauth[cor]{Corresponding author.}
\ead{Kai.Gallmeister@theo.physik.uni-giessen.de}
\author{T.~Falter}
\address{Institut f\"ur Theoretische Physik, Universit\"at Giessen, Germany}

\begin{abstract}
  We examine the space-time evolution of (pre-)hadron production
  within the Lund string fragmentation model. 
  The complete four-dimensional information of the string breaking
  vertices and the meeting points of the prehadron constituents are
  extracted for each single event in Monte Carlo simulations using the
  \Jetset{}-part of \Pythia{}. 
  We discuss the implication on the deep inelastic lepton scattering
  experiments at HERMES as well as on observables in
  ultra-relativistic heavy ion collisions at RHIC, using \Pythia{}
  also for modeling the hard part of the interaction.  
\end{abstract}
\end{frontmatter}

The fragmentation of high energy quarks into color-neutral hadrons
is a subject of high current interest. So far one cannot describe the complete 
space-time picture of this process within QCD as the underlying theory. 
The reason is that the formation of hadrons involves momentum scales of 
only a few hundred MeV which rules out a purely perturbative approach. 
Furthermore, the complexity of the fragmentation process goes far beyond 
what can be addressed by current non-perturbative lattice-QCD calculations. 

Estimates based on perturbative QCD \cite{DokBuch} suggest, that the
formation time of a hadron, i.e.~the time needed to build up the 
hadronic wave function, is of the order of 1\fmc{} in the hadron's restframe.
At large enough energies these eigentimes correspond to formation
lengths in the lab frame which are comparable with nuclear dimensions. 
If the fragmentation process is modified by the surrounding nuclear medium, 
an experimental comparison of high energy hadron production on various
nuclear targets with different radii can give insight into the
space-time picture of hadronization. With that respect the nucleus 
can be considered as a micro-laboratory for fragmentation studies.

In the recent past the HERMES collaboration at DESY has 
carried out an extensive experimental study of hadron attenuation 
\cite{HERMES-Exp} in deep inelastic lepton-nucleus scattering (DIS). 
Depending on the evolution of the fragmentation process in space-time 
the observed signals might either be dominated 
by (pre-)hadronic final state interactions (FSI) \cite{Acc,Kop,Fal,Falter} or by a
partonic energy loss prior to fragmentation \cite{WangArl}. A detailed
investigation of this subject is also essential for the interpretation of 
ongoing experiments at hadron colliders like RHIC or the future LHC which 
among other topics intend to find experimental evidence for a new
QCD-phase, the Quark-Gluon-Plasma (QGP). 
(Pre-)hadronic FSI could contribute to the suppression of
high-$p_\perp$ hadrons and thereby modify  jet-like signatures that
are still considered clean proofs for the creation  
of a deconfined QGP phase \cite{Gall,HSDKai}.

The Monte Carlo event generator \Pythia{} \cite{Pythia} has become one 
of the standard tools to perform calculations in the kinematic
regimes that we have just mentioned \cite{Falter,Gall,Patty}. 
The fragmentation in \Pythia{} is based on the Lund string model
\cite{Lund}. 
In former versions of \Pythia{} the routines handling the string
fragmentation were bundled in a separate package called \Jetset{}.
We keep this (old fashioned) nomenclature throughout this work in order
to  distinguish between the part of the event generator that creates
the hadronic strings (e.g.~\Pythia{}, \Fritiof{} \cite{FRITIOF} etc.) and 
the part that is responsible for the string fragmentation (\Jetset{}). 
A detailed description of the underlying model of \Jetset{} can be found
in Ref.~\cite{Sjo84}. In \Jetset{} the eigentimes of the string 
breaking vertices are determined during the fragmentation. 
However, the real essential information, namely the 
four-dimensional space-time coordinates of the string breaking, is
not reported. 

In this work we present a detailed quantitative analysis of the
four-dimensional space-time picture of hadron production with \Pythia{}/\Jetset{} 
in soft and hard interactions at HERMES and RHIC energies. We show for the first time
the exact spatiotemporal distribution of (pre-)hadron production and
formation points in realistic $ep$ and $pp$ collisions at HERMES and RHIC. Thereby we take into
account complications like resolved photon interactions, gluon radiation, primordial transverse momenta of quarks,
transverse momentum generation in the string fragmentation, cluster decay, finite mass effects, quark flavor dependence, diquark production, etc. Although our results should not be overstressed since one applies a semi-classical picture to a quantum mechanical problem,
our approach makes it possible to assign the four-dimensional production and formation point to each single hadron in each single scattering event.
This sets the foundation for consistently using the \Pythia{} model for both the determination of hadron momentum spectra -- where the model has been successfully tested -- and the description of the space-time evolution of hadron production. Both information can then be used in a consistent way as input for future transport simulations such as the BUU \cite{Fal} or the HSD model \cite{HSDKai} to describe nuclear reactions at HERMES and RHIC energies
and to determine the contribution of the (pre-)hadronic FSI to the experimentally observed hadron attenuation. Since the application of our findings is not restricted to one single nuclear reaction model, we exclude any specific modeling of the (pre-)hadronic FSI at this stage and merely concentrate on the space-time evolution of the (vacuum) fragmentation process in this work.

Although we use \Pythia{} to produce the initial string configurations we emphasize, that this is no restriction concerning the
extraction of the four-dimensional fragmentation coordinates, since
the fragmentation routines are independent of the string-building
routines. It is therefore ``ab initio'' straightforward, to transfer
our method to other event generators, supposed they use \Jetset{} for
fragmentation.

In principle, one could distinguish two classes of `strings': strings
that are made up of a quark and an antiquark only and those with
additional (hard) gluons located between the string ends as
transversal excitations.
(We will not touch strings build up entirely of gluons in this work.
Here the mechanisms at work are far beyond being transparent.
On the other hand, they are very rarely produced at the HERMES
energies and also play a minor role interpreting current RHIC findings.) 
However, this distinction is purely artificial, because the latter
reduces to the first case in the limit of vanishing transverse gluon
momenta \cite{Sjo84}. 
We point out that \Pythia{} also allows for diquarks at the string
ends as well as the creation of diquark-antidiquark pairs in the
string fragmentation.  
This possibility is accounted for in our computer code but omitted for
the sake of better readability in the following text. 

\Jetset{} includes two different hadronization methods for an initial
partonic system: 'string fragmentation' and 'cluster decay' where only 
the first one is based on the actual Lund fragmentation model. The latter 
method is constructed in such a way that it reproduces the rapidity 
distributions of the Lund fragmentation. Which of these two methods is 
used for hadronization depends mainly on the invariant mass of the 
decaying system.

\begin{figure}[tb]
  \begin{center}
    \hspace*{\fill}
    \includegraphics[width=8cm]{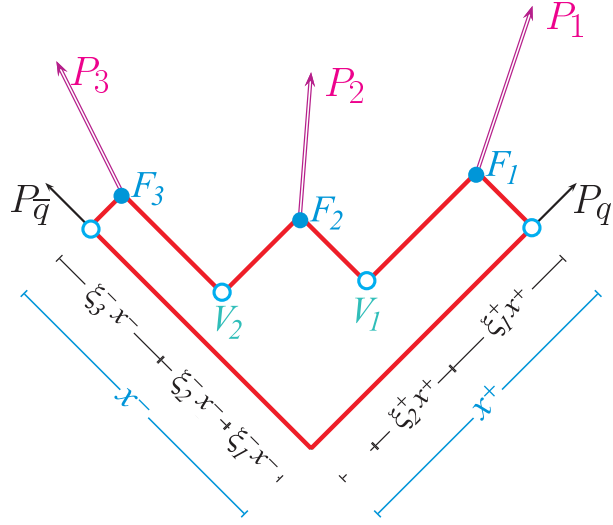}
    \hspace*{\fill}
    \caption
    {\textit{
        A sketch of the fragmentation of an one-dimensional 
        $q\bar q$-String (in the rest frame of the string). 
        The $P_x$ denote momenta, while $V_i$
        indicate string breaking vertices (production points) and 
        the $F_i$ the yoyo-formation points of the hadrons.
        While in principle all vectors are four-dimensional, in this
        sketch the horizontal axis is the $x$-axis and the vertical
        coordinate is the time $t$.
        Please note: not lightcone formalism, but four-vectors.
        }}
    \label{fig:qqString}
  \end{center}
\end{figure}

We begin with a review of the fragmentation of an
one-dimensional string, i.e.~a string that consists of a quark and
an antiquark string end and no gluonic excitation. Note that, most 
authors that discuss 'string fragmentation' restrict themselves to this 
one type of all possibilities. It can easily be illustrated and analytic 
expressions can be found, e.g.~in Refs.~\cite{Acc,Lund,Bialas}. 
We thus explain the necessary formalism by means 
of this example (cf.~Fig.~\ref{fig:qqString}). We stress, that 
all momenta and coordinates in Fig.~\ref{fig:qqString} 
are four-vectors and not light-cone variables.

In the following, we set the string tension to $\kappa=1\GeV/\fm$. 
This allows us to omit $\kappa$ in all
formulae and to identify momentum vectors with space-time coordinates,
i.e.
\begin{equation}
\label{eq:qqString}
P_q=x^+\,,\quad P_{\bar q}=x^-
\end{equation}
where $P_q$ ($P_{\bar q}$) and $x^+$ ($x^-$) denote the initial 
four-momentum and turning-point of the (anti-)quark respectively.
The four-momenta $P_i$ of the final hadrons are given by
\begin{equation}
\label{eq:qqString_P}
P_i = \xi_i^+x^++\xi_i^-x^-
\end{equation}
with the constraints $0\leq\xi_i^\pm\leq 1$ and $\sum\xi_i^\pm =1$. 
For each hadron $\xi_i^+$ and $\xi_i^-$ are obviously related via 
the (tranverse) mass $m_i$ of the produced hadron. 
In Eq.~(\ref{eq:qqString_P}) we have neglected any transverse momenta 
created in the string breakings for simplicity. The following formulae are 
not affected by this simplification.
The space-time positions of the string breakings (production points) are
\begin{equation}
\label{eq:qqString_V}
V_i = \chi_i^+x^++\chi_i^-x^-  
\end{equation}
with $\chi_i^+=1-\sum_{j=1}^i\xi_j^+$, $\chi_i^-=0+\sum_{j=1}^i\xi_j^-$
and the points where the world-lines of the constituents of 
hadron $i$ meet for the first time -- usually called the 
'yoyo-formation point' of hadron $i$ -- are given as
\begin{equation}
\label{eq:qqString_F}
F_i = \chi_i^+x^++\chi_{i+1}^-x^-    \ .
\end{equation}
One should keep in mind, that the latter only makes sense if one
neglects any transverse momentum of the $\bar q q$-pairs created at
the string breaking vertices. In general, a quark from one vertex
-- having some transverse momentum -- does not cross the trajectory
of an antiquark that stems from another vertex and has some other
transverse momentum. We point out that the yoyo-formation points
do not necessarily correspond to the real formation points of the 
hadron, i.e.~the space-time point where the full hadronic 
wavefunction has built up. 

Obviously, both the production and formation points in space-time can be 
reconstructed from the initial quark and antiquark momenta $P_q$ and 
$P_{\bar q}$ if one extracts the complete set of $\xi_i^\pm$ 
from \Jetset{}. (Please recall: All the transversal momentum
components are only necessary to reconstruct the hadron momenta, not
the fragmentation points.)

We now concentrate on the implementation of the
string fragmentation in \Jetset{}. Like all
other Lund-implementations it suffers from the fact that
in the derivation of the fragmentation function one 
assumes that the decaying string has infinite 
mass \cite{Sjo84}. 
If one starts on one side of the string 
(e.g.~the quark string end) and continues the fragmentation 
to the other side (i.e.~the antiquark string end), 
one will run into the problem of finding a solution for the 
last hadron that both conserves 
energy and momentum and fulfills the mass-shell condition
for the hadron. In order to avoid this problem, 
\Jetset{} randomly performs fragmentations at one or the other side
of the decaying string, 
until the invariant mass of the remaining string system 
(somewhere in the middle of the string) falls below some 
threshold value. Then \Jetset{} switches to a mode in which the 
parameters of the two final hadrons are chosen
simultaneously in such a way that energy and momentum are
conserved and that the rapidity distribution of the hadrons
looks like in the original Lund fragmentation process 
\cite{Pythia,Ede}. This method is very
similar to the one used for the 'cluster decay' discussed below.

So far, the production and formation points could be determined by 
Eqs.~(\ref{eq:qqString_V}) and (\ref{eq:qqString_F}) respectively.
However, strings may have transverse excitations,
which are represented by additional gluons between the string ends. 
These gluons impose some 
transverse momentum on the string and lead to a complex movement of 
the string in space-time which is known as the 'dance of the butterfly' 
\cite{Lund}. 

The additional gluons split the string into separate regions. 
For a configuration that consists of a quark,
$n-2$ gluons and an antiquark with related four-momenta 
$q(p_1)$, $g(p_2)$, $\ldots$ , $g(p_{n-1})$, $\bar q(p_n)$ the initial 
string contains $n-1$ pieces. In four-momentum space the $i$-th string 
piece is spanned by the two four-momenta $p^+_{(i)}$ and $p^-_{(i)}$ 
where we have defined
$p^+_{(1)}=p_1$, $p^-_{(1)}=p_2/2$, $p^+_{(2)}=p_2/2$, 
$p^-_{(2)}=p_{3}/2$, $\ldots$ , $p^+_{(n-1)}=p_{n-1}/2$, $p^-_{(n-1)}=p_n$
in analogy to the simple $q\bar q$-string (Eq.~(\ref{eq:qqString})). The factors
$1/2$ arise because the gluons share their energy between the two adjacent
string pieces.

Due to the complex movement of the string new regions appear 
as time goes by. Each region can be thought of being spanned by one 
$p^+_{(j)}$ and another $p^-_{(k)}$ four-vector with $j$ and $k$
not necessarily adjacent. It is straightforward, but tedious and not
illuminating, to extend 
Eqs.~(\ref{eq:qqString_P}) and (\ref{eq:qqString_F}) to the case of
multiple string pieces \cite{Sjo84}.
Therefore we only comment on the major complications: 
Since now the string(-part) ends have initial ``transversal'' momentum
components, transverse momenta of the quarks originating from the
string breakings must be considered explicitly. 
Hadrons built up from partons that are created in
different string regions cause severe problems in the string 
fragmentation. Simplifications that are made in the \Jetset{} 
code to cure this problems result in faulty four-dimensional 
coordinates of the fragmentation points  \cite{Sjo84}.

As pointed out not every string configuration in \Jetset{} decays 
according to the 
Lund fragmentation scheme. Depending on the invariant mass of the
system it may also fragment according to a cluster decay algorithm 
where the final state may only consist of one or two hadrons.
In case of the decay into one hadron, the strings's 
four-momentum is mapped onto the four-momentum of the final 
hadron. Since the hadron mass is more or less fixed, left over 
components are transfered to a nearby string in form of a gluon with 
the corresponding four-momentum.
If the cluster decays into two hadrons, the same algorithm as for 
the two last hadrons in the usual string fragmentation is applied.
For details we refer the reader to Ref.~\cite{Pythia} and references
therein.
Production vertices of hadrons stemming from cluster decays are
assumed to be at the interaction point. In case of a cluster decay into
two hadrons, the four-dimensional information of the second, 
i.e.~the ``middle'' quark-antiquark
production point, is reconstructed from energy-momentum constraints
in analogy to Eq.~(\ref{eq:qqString_V}). Also the formation points
are determined from the four-momenta of the hadrons in analogy to 
Eq.~(\ref{eq:qqString_F}).

Since \Jetset{} does not report any space-time vectors concerning the
fragmentation progress, this has to be done manually. We enhanced
the corresponding routines to report some informations such as the 
$\xi_i^\pm$, which enable us to calculate each string breaking point 
and all meeting points in four dimensions. Please recall, that this is 
all the information we need.
We have checked, that the proper times reconstructed from these values
match those used in the fragmentation process.

\begin{figure}[tb]
  \begin{center}
    \hspace*{\fill}
    \includegraphics[width=11cm]{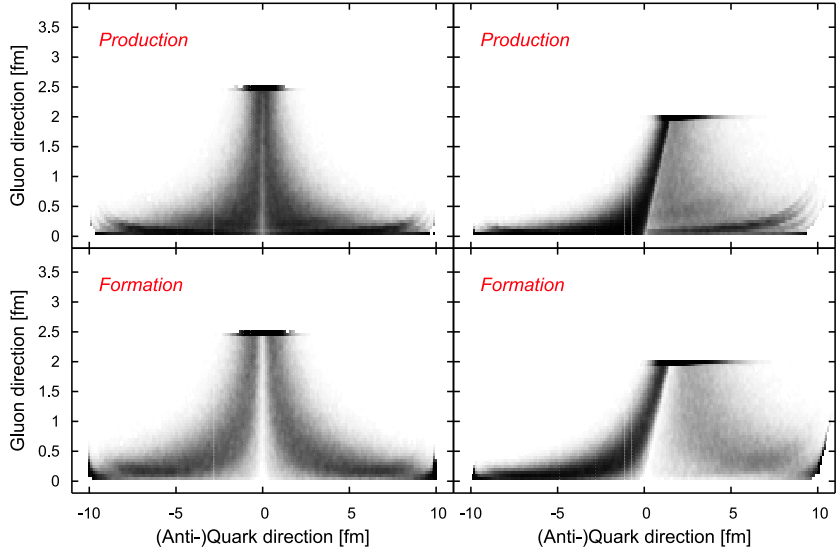}
    \hspace*{\fill}
    \caption
    {\textit{Distribution of production (upper panel) and formation 
        points (lower panel) for the example $P_{q,\bar q}=(\pm
        10,\,0)\GeV,\ P_g=(0,\,5)\GeV$ (left) and $P_g=(3,\,4)\GeV$
        (right). Dark regions refer to larger decay and formation
        probabilities.  
        }}
    \label{fig:qgq_Perp}
    \label{fig:qgq_Tilt}
  \end{center}
\end{figure}

As an example we consider the fragmentation of a $qg\bar q$-string
with \Jetset{}.
For simplification purposes we switched off transverse momentum production
at the string breaking vertices.
On the left hand side of Fig.~\ref{fig:qgq_Perp} we show the 
spatial distributions of the string breakings, 
i.e.~production points, and the 
yoyo-formation points for a system where the quark and antiquark 
momenta ($\pm 10\GeV$) are along the $x$-axis 
(with opposite directions) and the gluon moves
perpendicular to them into the $y$-direction ($5\GeV$). 
The spatial boundaries of the production points are
set by the initial (anti-)quark and gluon momenta. Since the gluon
looses energy to both the adjacent string pieces the production
points have transverse coordinates $y\leq 2.5$ \fm. As one might 
have expected, the yoyo-formation points are slightly shifted outwards 
compared to the production points.
Taking the same parameters, but slightly tilting the gluon momentum out
of the transversal direction, the spatial distributions change according
to the right hand side of Fig.~\ref{fig:qgq_Tilt}.
Please note however, that these are more or less pathological
examples, since in real event simulations (some hard scattering with
initial and final state radiation) string regions with invariant
masses as large as chosen here are quite rare.

The sharp structures, best seen in the distribution of the production
points for the tilted gluon (Fig.~\ref{fig:qgq_Tilt}, right top), are
due to mass constraints and only visible, if one neglects transversal
momenta, as done in this example: Structures caused by different hadron 
masses $m$ are smeared if one has to switch to the transverse mass $m_\perp$.
We point out, that we have excluded production and formation
points, which are due to a faulty reconstruction caused by the abovementioned
simplifications in the \Jetset{} code. These would give rise to spurious
points at transversal components larger than what is allowed by the gluon 
component and contribute with less than two percent to the total yield. 

In both cases one clearly identifies a new string region
as the horizontal distribution displaced from the $x$-axis.
This region appears in the time-evolution of the string
after the gluon has lost all of its momentum to the string.

We now apply our method of extracting four-dimensional vertices 
to two practical examples. The first is an electron (virtual gamma) 
induced reaction on a nucleon at HERMES energies and the
second one is a $pp$-collision at RHIC energies. 

The experimental data on $\gamma^*p$ and $\gamma^*A$ reactions at HERMES
imply some complicated kinematic cuts, e.g.~in the $\nu$-$Q^2$-plane where
$\nu$ denotes the photon's energy and $Q^2$ its virtuality. 
In this study we will neglect this complication and simply
replace the whole averaging over a multi dimensional parameter space
by using some mean values. Since the underlying physical processes vary
drastically (e.g.~importance of resolved photon-nucleon interactions) 
within the full experimentally accessible kinematical region, 
our example is not representative for {\it all} possible HERMES events.
\begin{figure}[tb]
  \begin{center}
    \hspace*{\fill}
    \includegraphics[width=11cm,angle=-90]{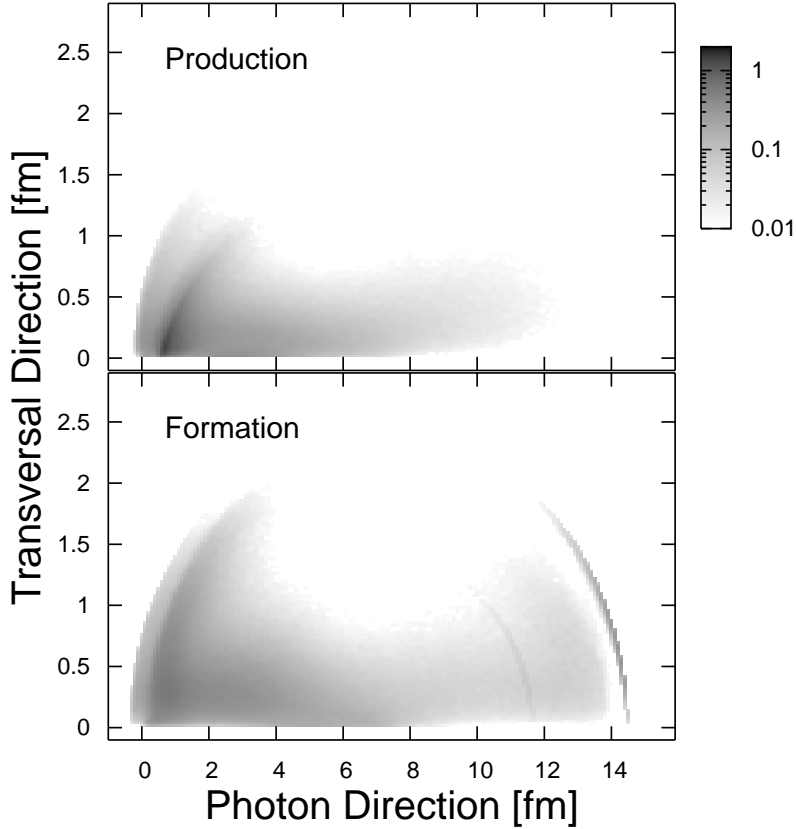}
    \hspace*{\fill}
    \caption
    {\textit{Production (top) and formation (bottom) points for
        a typical Hermes event: $(\nu=14\GeV,\
        Q^2=2.5\GeV^2)$. The target nucleon is located at the origin, 
        the virtual photon is coming in from the left. Labels in the
        plots correspond to explanations in the text.
        }}
    \label{fig:Hermes}
  \end{center}
\end{figure}
In Fig.~\ref{fig:Hermes} we show the spatial distribution of production
and formation points for a $\gamma^*p$-reaction at 
$\langle\nu\rangle=14\GeV$ and $\langle Q^2\rangle=2.5\GeV^2$. 
This corresponds to an 
invariant mass $\langle W\rangle = 5\GeV$ of the photon-nucleon system.
The photon is coming in from the left and strikes a nucleon which is
located at the origin. While PYTHIA describes all spectra of
non-charmed hadrons at HERMES energies astonishingly well 
\cite{Fal,Falter,Patty}, its ability to describe charm production 
has never been tested in this kinematic regime. We therefore exclude 
all charmed hadrons from our analysis. Note that both production and 
formation points lie mainly inside a radius of $\sim6\fm$ around the
interaction point. If one compares these values with a typical diameter 
of a complex nucleus one might expect strong (pre-)hadronic FSI after the 
(production) formation time in case of photonuclear reactions. 

In contrast to common belief the DIS does not simply lead
to a one-dimensional string configuration. Instead of that, the 
distributions in Fig.~\ref{fig:Hermes} show some (elliptical)
structures which are again due to non vanishing mass terms and 
need some more detailed explanations:
Due to kinematics all production and formations points lie within
circles of radius $W/2$ in the cm-frame of the photon-nucleon
system. By boosting to the lab frame one obtains the observed ellipses with
maximum transverse dimension $W/2$ and extremal longitudinal values of
$x_z^-\simeq-0.4\fm$ and $x_z^+\simeq 14.5\fm$. 

The transversal distribution of the vertices is caused by two mechanisms. 
The first one is the transverse momentum $p_\perp$
generated in the string breakings that leads to a small transverse 
spread which is best seen at longitudinal distances around  $7\fm$.
The by far more important effect, however, stems from the intrinsic transverse
momentum $k_\perp$ of the nucleon constituents. This can easily
be understood by considering the simple scenario, where the virtual
photon gets absorbed by a single quark inside the proton. In that case
a single string is spanned between a diquark at the origin and the struck 
quark. Without intrinsic momenta, such a string would simply expand along the 
initial photon direction. The intrinsic $k_\perp$ rotates the direction
of the string (without changing its invariant mass) and leads to the transversal 
distribution of production and formation points. 
(We used the default value $\langle k_\perp^2\rangle = 1\GeV^2$ of \Pythia{} 
in our simulations.)

\begin{figure}[tb]
  \begin{center}
    \hspace*{\fill}
    \includegraphics[width=7cm,angle=-90]{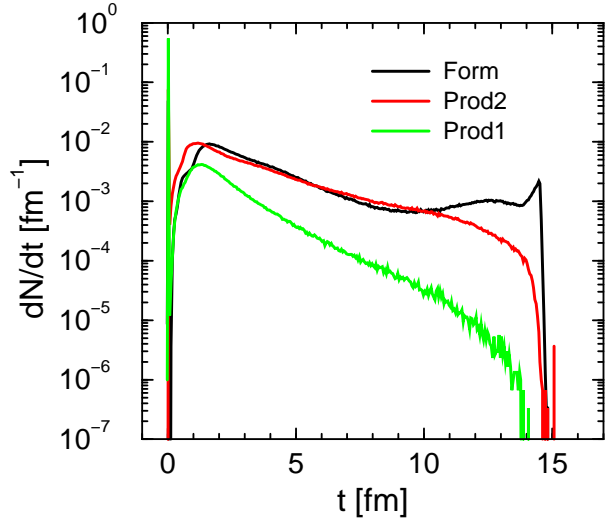}
    \hspace*{\fill}
    \caption
    {\textit{Distributions of the production times $t_1$, $t_2$ and 
    the formation time $t_F$ for a virtual $\gamma^*p$ event. The kinematics
    are the same as for Fig.~\ref{fig:Hermes} $(\nu=14\GeV,\ Q^2=2.5\GeV^2)$.}}
    \label{fig:dNdt}
  \end{center}
\end{figure}
The yoyo-formation point of the first-rank hadron, i.e.~the hadron labeled '1' 
in Fig.~\ref{fig:qqString}, is given by $x^++\xi_1^-x^-$. In the case of one single
quark-diquark string its longitudinal coordinate in the lab frame is therefore confined
to the narrow region between $14.1\fm$ and $14.5\fm$ if one neglects all transverse 
momenta. Correspondingly, the formation point of the highest-rank hadron, i.e.~the hadron 
labeled '3' in Fig.~\ref{fig:qqString}, is given by $x^-+\xi_n^+x^+$ and, hence, its 
longitudinal coordinate lies between $-0.4\fm$ and $14.1\fm$. 
The finite invariant mass $W$ of the string gives rise to a lower boundary 
($\xi^\pm\geq(m_h/W)^2$) for the possible $\xi$-values. This lower limit
leads to the two distinct branches that are visible in Fig.~\ref{fig:Hermes}: While
the lower limit is negligible for pions, it becomes important for 
the heavier mesons and baryons; as a consequence the latter have formation points 
with larger (here: positive) longitudinal component.
A similar (but more involved) mechanism is at work for the
production points and leads to the strong (dark) band at low
longitudinal positions. 

The importance of the production points for hadron attenuation in nuclear reactions
has been extensively discussed in Refs.~\cite{Acc,Kop,Fal,Falter,HSDKai}.
The constituents of a hadron that is formed at point $F_i$ in Fig.~\ref{fig:qqString} 
are produced at the string-breaking vertices $V_i$ and $V_{i-1}$. We denote the larger
of the two production times $t_2$ and the smaller one $t_1$. At the later production 
point, i.e.~at time $t_2$, the string splits into a color neutral object (prehadron) 
-- which has the four-momentum of the final hadron -- and a remainder string. 
The final state interactions of this prehadron with the 
nuclear environment modify the spectrum of the ultimately detected hadrons but do not 
influence the actual string fragmentation process. Fig.~\ref{fig:dNdt} shows the distributions 
of the production times $t_1$ and $t_2$ of the first and second hadron constituents as well as the 
yoyo-formation times $t_F$. The structures in the time distributions simply reflect the
complex features that already showed up in Fig.~\ref{fig:Hermes}. The presented
distributions might be used as input for various attenuation studies such as the 
recent prehadron absorption model of Ref.~\cite{Grunewald}.

\begin{figure}[tb]
  \begin{center}
    \hspace*{\fill}
    \includegraphics[width=7cm,angle=-90]{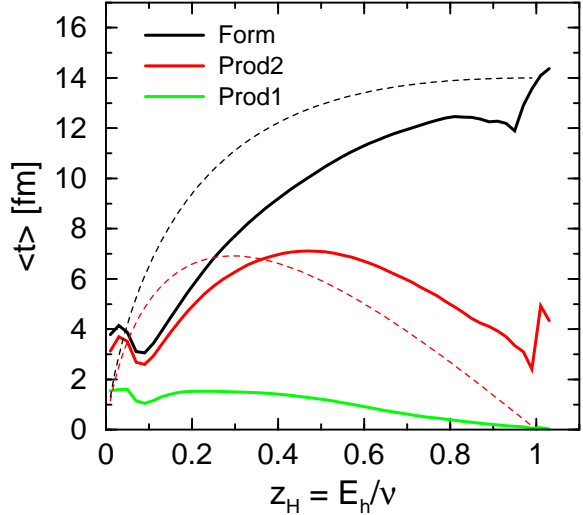}
    \hspace*{\fill}
    \caption
    {\textit{The solid curves show the average production and 
    formation times of hadrons in certain bins of the fractional 
    hadron energy $z_h=E_h/\nu$. The kinematics
    are the same as for Fig.~\ref{fig:Hermes}
    $(\nu=14\GeV,\ Q^2=2.5\GeV^2)$.
    The dashed curves represent the results of the analytic 
    estimate derived in Ref.~\cite{Bialas} for the decay of an 
    one-dimensional string into infinitely many fragments.}}
    \label{fig:zh-times}
  \end{center}
\end{figure}
In Fig.~\ref{fig:zh-times} we show the average production times
$\langle t_{1}\rangle$ and $\langle t_{2}\rangle$ as well as the 
average formation time
$\langle t_F\rangle$ as function of the fractional hadron energy
$z_h=E_h/\nu$. One clearly sees that the hadron formation times 
increase with energy whereas the production times of prehadrons become 
small if they carry a large energy fraction $z_h$. 
According to the Lund model the prehadronic 
final state interactions therefore dominate the attenuation of high energy hadrons
in DIS of nuclei.

For comparison Fig.~\ref{fig:zh-times} also shows the simple time estimates of 
Ref.~\cite{Bialas} that were derived analytically for the fragmentation of an 
one-dimensional string into infinitely many fragments. Obviously, 
the analytic estimates yield only a rough approximation of the average production 
and formation times in \Jetset{}. Furthermore, it is not enough to know 
the average production and formation times
if one wants to perform realistic calculations of nuclear reactions. 
Our method allows for a complete event-by-event simulation 
assigning the production and formation points and times to all hadrons
in each \Jetset{} event.

\begin{figure}[tb]
  \begin{center}
    \hspace*{\fill}
    \includegraphics[width=7cm,angle=-90]{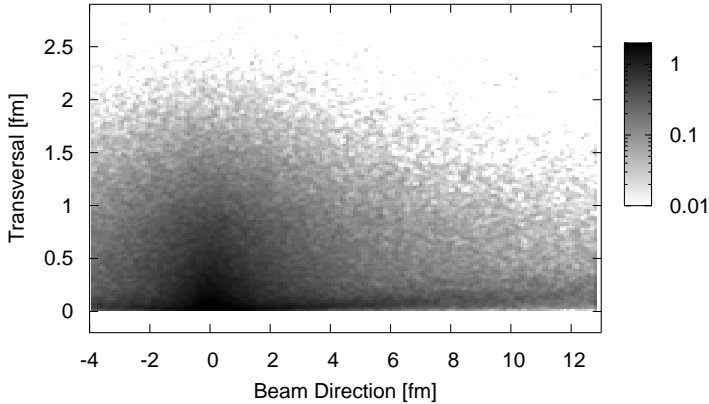}
    \hspace*{\fill}
    \caption
    {\textit{Production points for RHIC: $N+N$@$\sqrt s=200\GeV$. The
        picture is symmetric in horizontal direction, the contribution
        extends up to $\pm 100\fm$.
        }}
    \label{fig:RHIC_allY}
  \end{center}
\end{figure}

The distribution of production and formation points for $pp$-collisions at
RHIC ($\sqrt{s}=200\GeV$) shown in Fig.~\ref{fig:RHIC_allY}
can be easily understood: Around the interaction point -- corresponding 
to mid-rapidity -- one has the broadest transverse momentum distribution
and hence the largest transverse coordinates of the 
production and formation points. 
The regions with rapidity $y\to\infty$ mainly consist of remnants, 
that lead to hadron production along the beam axis. 
Since the rate of particles $dN/d\pT dy$ decreases approximately
exponentially with increasing $\pT$, also the distribution of
production and formation points in transversal direction does.
(This is in contrast to the HERMES regime, where the photon sets an 
upper limit on the energy of the particles. Here this upper limit far out 
of reach and the transverse distribution is practically unconstrained.)

\begin{figure}[tb]
  \begin{center}
    \hspace*{\fill}
    \includegraphics[width=10cm]{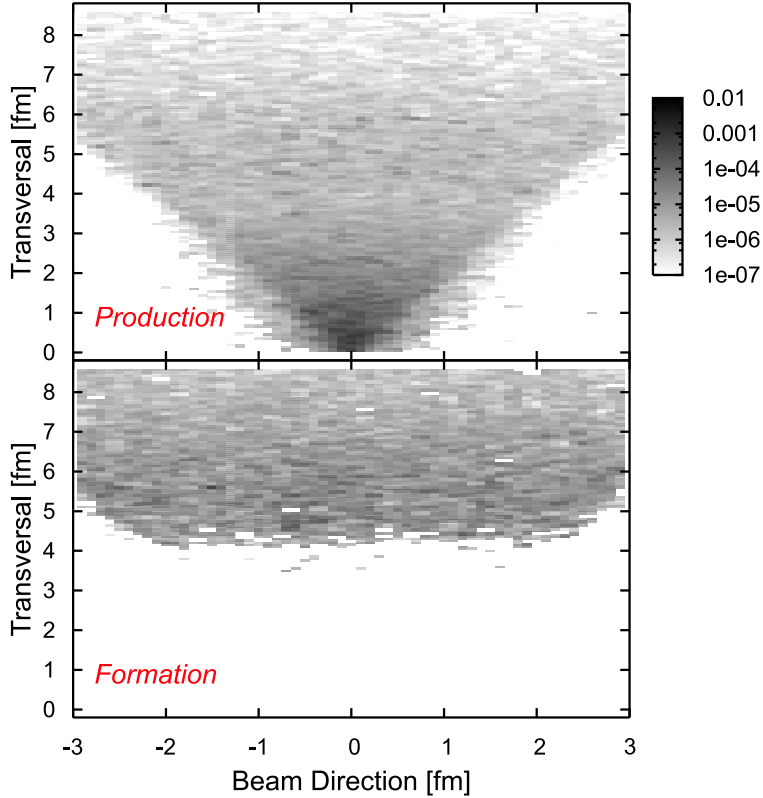}
    \hspace*{\fill}
    \caption
    {\textit{Production and formation points for RHIC($N+N$@$\sqrt s=200\GeV$) for
        mid-rapidity ($|y|<0.5$) and high $\pT{}$: $\pT>4\GeV$. 
        }}
    \label{fig:RHIC_MidY}
  \end{center}
\end{figure}

Of major experimental interest are high $\pT$-hadrons
(e.g.~$\pT>4\GeV$) at mid-rapidity ($|y|<0.5$).
The production and formation points applying these 
cuts are shown in Fig.~\ref{fig:RHIC_MidY}.
While the distribution of formation points supports the
popular belief that high-$p_\perp$ mid-rapidity
hadrons at RHIC are produced outside any fireball or deconfined 
phase, the distribution of production points tells another story: 
Most of the production points are within a transversal distance below $2\fm$.
However, we point out that the high hadron densities created
in ultra-relativistic heavy ion collisions might strongly influence the string 
fragmentation process and that in the presence of a deconfined phase (QGP) the
whole concept of hadronic strings becomes meaningless. For heavy ion collisions the presented production times and lengths should therefore be rather seen as lower estimates.

In order to quantify the findings from above, Tab.~\ref{tab:numbs}
presents the average proper times $\langle\tau\rangle$ of the
production and formation vertices as well as the average production
and formation times $\langle t\rangle$ in the laboratory frame. 
For comparison we also list the average values of the fraction $t/\tau$ 
and the Lorentz boost factor $\gamma=E_h/m_h$ of the considered hadron. 
Note that for each hadron there exist 
two production (proper-)times ($\tau_1$)
$t_1$ and ($\tau_2$) $t_2$ -- corresponding to the 
production of the first and second hadron constituent --
as well as one yoyo-formation (proper-)time ($\tau_F$) $t_F$.
The larger of the two production times $t_2$ is also called
prehadron production time \cite{Kop,Fal}.
\begin{table}[htb]
  \begin{center}
    \begin{tabular}{|c||c|c|c||c|c|c||c|c|c||c|}
      \hline
      & $\langle\tau_1\rangle$ 
      & $\langle\tau_2\rangle$ 
      & $\langle\tau_F\rangle$ 
      & $\langle t_1\rangle$ 
      & $\langle t_2\rangle$ 
      & $\langle t_F\rangle$
      & $\langle \frac{t_1}{\tau_1}\rangle$ 
      & $\langle \frac{t_2}{\tau_2}\rangle$ 
      & $\langle \frac{t_F}{\tau_F}\rangle$
      & $\langle \gamma \rangle$ \\
      & \fm 
      & \fm 
      & \fm 
      & \fm 
      & \fm 
      & \fm
      & 
      & 
      & 
      & \\
      \hline
      \multicolumn{11}{|c|}{HERMES: $W=5\GeV$, $Q^2=2.5\GeV^2$} \\
      \hline
      $\pi$  & 0.24& 0.81& 0.92& 0.83& 3.49& 4.37& 3.8&5.3&6.2&13.53\\
      $\rho$ & 0.21& 0.95& 1.27& 0.74& 3.28& 4.76& 3.9&3.6&4.0& 5.60\\
      $N$    & 0.18& 0.74& 1.11& 0.42& 1.62& 2.11& 2.5&2.2&1.8& 2.04\\
      \hline
      \multicolumn{11}{|c|}{RHIC: $|y|<0.5$, $p_\perp>4\GeV$} \\
      \hline
      $\pi$  & 0.36& 0.81& 1.03& 2.34& 8.19& 10.30& 7.1&13.8&18.7&51.04\\
      $\rho$ & 0.38& 0.99& 1.42& 2.28& 6.62& 10.87& 7.0&6.9&8.5&9.31\\
      $N$    & 0.62& 1.33& 1.82& 3.10& 7.55& 10.61& 5.7&6.0&6.4&7.12\\
      \hline
    \end{tabular}

    \caption
    {\textit{Average production and formation times for pions, $\rho$
     mesons and
      baryons within specified kinematics (``HERMES'',
      ``RHIC''). Proper times are labeled by $\tau$, 
      while $t$ represents the corresponding time in the laboratory frame.
      $\gamma$ is the relativistic boost factor of the hadron.}}
     \label{tab:numbs}
  \end{center}
\end{table}

The average production and formation {\it proper times} at HERMES seem 
to be almost independent of the particle species. 
On the other hand, one observes the same universal feature 
for the average {\it laboratory times} at RHIC conditions (with the implemented cuts).
Furthermore, the ratio $\langle t/\tau\rangle$ deviates from the 
Lorentz factor $\langle\gamma\rangle$ of the corresponding hadron.
Hence, the times that we have extracted from \Jetset{} are
in contradiction to the widely-used assumption that
the hadron production and formation times in the lab frame are simply 
given as a product of the Lorentz-boost factor $\gamma$ and
an universal constant proper time.

In the presented work we have extracted the production 
and formation vertices of hadrons from the Lund fragmentation
as modeled by \Jetset{}. 
In contrast to other approaches, such as analytic derivations, we are
able to extract the full four-dimensional information for each hadron
produced in the fragmentation of a (gluonic) string on an event-by-event 
basis. For the first time we have presented quantitative results for realistic 
photon-nucleon reactions at HERMES energies and $pp$-collisions at RHIC
using the Monte Carlo generator \Pythia{}. However, we stress that our
method can also be used for all other kinds of event generators which
are based on \Jetset{} for the fragmentation. 
We find that a large fraction of
production and formation points at HERMES energies lie inside
typical nuclear radii which underlines the importance of (pre-)
hadronic FSI. Also the production points of high-$p_\perp$ 
mid-rapidity hadrons at RHIC seem to lie inside the fireball
region. Furthermore, we conclude that the conventional determination 
of hadron production and formation times by simply multiplying 
a (universal) constant proper time with the corresponding 
$\gamma$ factor leads to misleading estimates for the
hadron production and formation points.

This work is thought to be a 'proof of concept'. Detailed
transport-theoretical studies of nuclear reactions \cite{Fal,HSDKai} 
that explicitly model the (pre-)hadronic FSI in the nuclear environment will follow. We point out that any event-by-event simulation based on our
findings can account for all sorts of experimental cuts and detector efficiencies. Such models can therefore be directly compared with experimental data or used for detector simulations. They will help to clarify to what extend the experimentally observed hadron attenuation in nuclear DIS and ultra-relativistic heavy ion collisions is caused by the rescattering of the produced colorneutral (pre-)hadrons after fragmentation and what is caused by additional in-medium phenomena like a modification of the fragmentation function, a partonic energy loss or recombination effects.

This work has been supported by BMBF.


\end{document}